\documentclass[
aps,
pra,
twocolumn,
supescriptaddress,
showpacs,
floatfix
]{revtex4-2}
\usepackage{adjustbox}
\usepackage{verbatim}
\usepackage[colorlinks=false,bookmarks,pdftitle={},pdfauthor={},pdfsubject={},pdfkeywords={},breaklinks]{hyperref}
\usepackage[usenames]{color}
\usepackage{amssymb,mathbbol,amsmath,amsbsy}
\usepackage{amsfonts}
\usepackage{dsfont}
\usepackage{amsmath}
\usepackage[utf8]{inputenc}
\usepackage{graphicx}
\usepackage{multirow}

\newcommand{\alphaS}{\alpha_\mathrm{S}}
\newcommand{\alphaA}{\alpha_\mathrm{A}}
\newcommand{\betaS}{\beta_\mathrm{S}}
\newcommand{\betaA}{\beta_\mathrm{A}}
\newcommand{\gammaS}{\gamma_\mathrm{S}}
\newcommand{\gammaA}{\gamma_\mathrm{A}}
\newcommand{\kappael}{\kappa_\mathrm{el}}
\newcommand{\kappaph}{\kappa_\mathrm{ph}}
\newcommand{\kB}{k_\mathrm{B}}
\newcommand{\zTmax}{zT_\mathrm{max}}

\begin{document}
\title{Dimensional crossover and enhanced thermoelectric efficiency\\ due to broken symmetry in graphene antidot lattices}
\author{M. Neşet Çınar}
\author{H. Sevin\c{c}li}
\email[Corresponding author: ]{haldunsevincli@iyte.edu.tr}
\affiliation{
Deparment of Materials Science and Engineering, \.Izmir Institute of Technology, G\"ulbah\c{c}e Kamp\"us\"u, 35430 Urla, \.Izmir, Turkey.}

\begin{abstract}
Graphene antidot lattices (GALs) are {two-dimensional (2D)} monolayers with periodically placed holes in otherwise pristine graphene.
We investigate the electronic properties of symmetric and asymmetric GAL structures having hexagonal holes, and show that anisotropic {2D} GALs can display a dimensional crossover such that {effectively} one-dimensional (1D) electronic structures can be realized in two-dimensions around the charge neutrality point. 
We investigate the transport and thermoelectric properties of these {2D} GALs by using nonequilibrium Green function (NEGF) method. 
Dimensional crossover manifests itself as transmission plateaus, a characteristic feature of 1D systems, and enhancement of thermoelectric efficiency, where thermoelectric figure of merit, $zT$, can be as high as 0.9 at room temperature.
We also study the transport properties in the presence of Anderson disorder and find that mean free paths of {effectively} 1D electrons of anisotropic configuration are much longer than their isotropic counterparts.
{We further argue that dimensional crossover due to broken symmetry and enhancement of thermoelectric efficiency can be nanostructuring strategy virtually for all 2D materials.}
\end{abstract}

\maketitle
\section{INTRODUCTION}
Graphene and related two-dimensional materials have revolutionized condensed-matter physics not only with the exceptional physical properties they possess, but also with the existence of seemingly infinite possibilities to tailor these physical properties through nanostructuring~\cite{geim2007,wu2010,dresselhaus2016,das2014,bai2010,moreno2018,Zong2020}.
Cutting two-dimensional materials into ribbons alters their properties drastically.
{Unlike semimetallic 2D graphene,} {one-dimensional} graphene ribbons are semiconducting~\cite{fujita1996,wakabayashi1999,son2006,yang2007} 
and can have substantially enhanced thermoelectric (TE) efficiencies~\cite{chen2010,sevincli:prb:2010,sevincli2013,xiao2018}. 
Another scheme for nanostructuring is creating holes.
GALs are {2D} structures with periodic arrays of holes~\cite{pedersen2008,furst2009,bai2010,sandner2015}.
They have a nonzero band gap, which is tunable with the geometrical parameters. 
Their thermal conductivity is also suppressed significantly~\cite{pedersen2008,furst2009,gunst2011,chang2012,yan:physletta:2012,dollfus2012,karamitaheri2011,sh2015}.

Enhancement of TE efficiency is an intricate problem because an efficient TE material requires to have a large Seebeck coefficient like in an insulator, high electrical conductivity like in a metal and poor thermal conduction like in a glass~\cite{slack95}.
{2D} materials have been subject of intense research in the last decade for thermoelectric applications~\cite{dollfus:jpcm:2015,chen:jpcc:2015,adessi:jpcc:2017,ozbal:prb:2019,li:mnl:2020,kanahashi:aem:2020}.
The  main motivation is the fact that reduced dimensionality enhances TE efficiency~\cite{hicks1993,hicks1993thermo}.
Another reason is the richness and power of nanostructuring possibilities.
Graphene is the most impressive example.
Pristine graphene's electronic structure acquire zero band gap and electron-hole symmetry while its phononic structure yields the record thermal-conductivity value~\cite{neto2009,balandin2008}.
A material with such properties is among the least likely ones to have good TE efficiency.
Still a number of nanostructuring schemes have proven that graphene could acquire extraordinary TE efficiencies~\cite{karamitaheri2011,sevincli:prb:2010,gunst2011,chang2012,sevincli2013,chang2014,nyugen2014,dollfus2015,rocha:chapter:2011,xu2014,Tran2017}.

\begin{figure}[b]
	\flushleft
	(a)\\
	\vspace{-4mm}
	\centering
	\includegraphics[width=7.0cm]{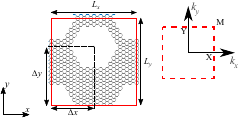}\\
	\vspace{3mm}
	\flushleft
	(b)\\
	\vspace{-7mm}
	\flushright
	\includegraphics[width=8.0cm]{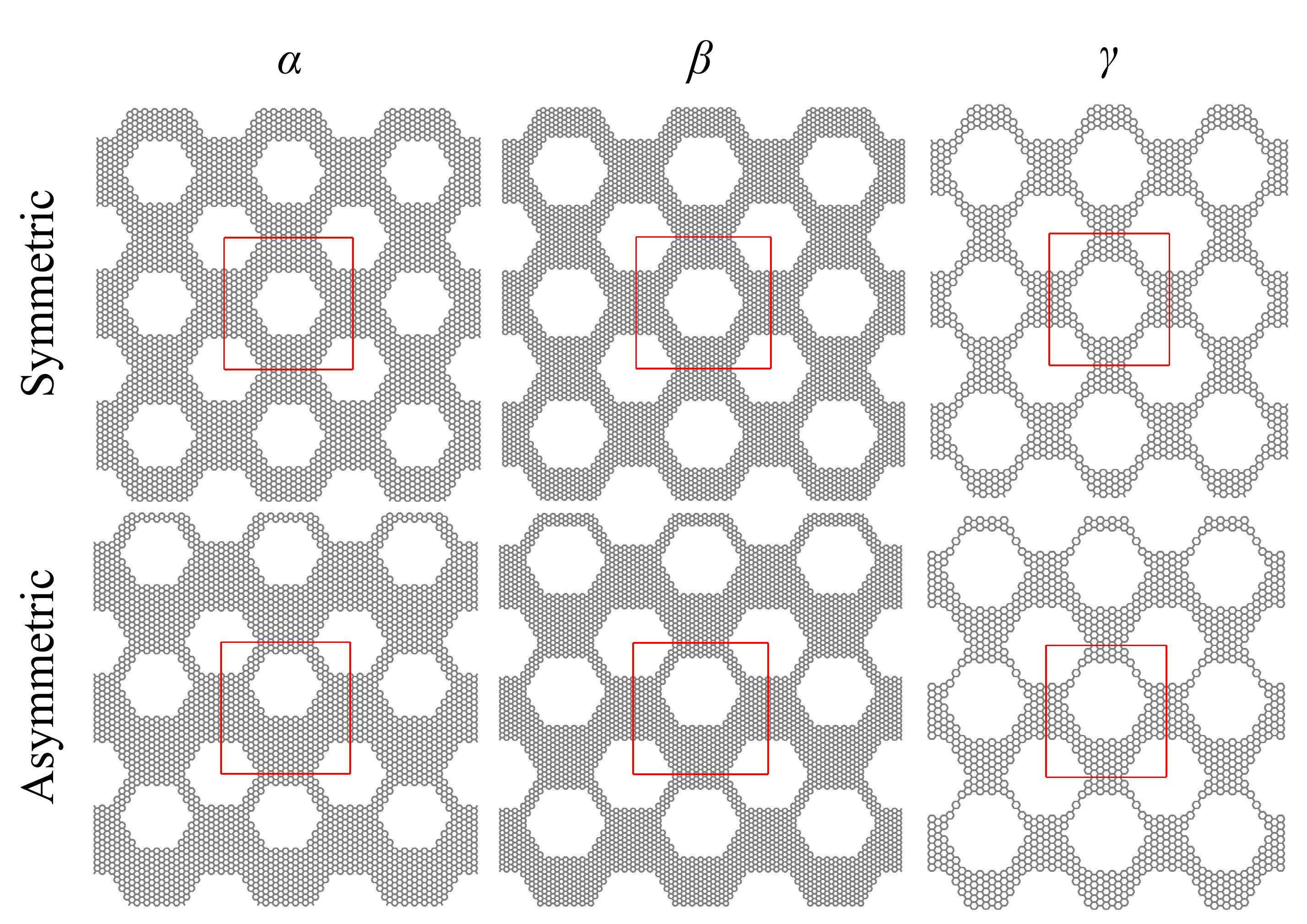}
	\vspace{-2mm}
	\caption{Structural parameters of GALs are shown in (a). 
		There are two antidots per unit cell, which has side lengths $L_x$ and $L_y$. Antidots are hexagonal and equal in size ($r$).
		The center of an antidot is chosen at the origin, while the second is at ($\Delta x$,$\Delta y$).
		The studied geometries ($\alpha$, $\beta$, $\gamma$) are shown below.
	}
	\label{fig:structure}
\end{figure}

\begin{figure*}
	\centering
	\includegraphics[width=18cm]{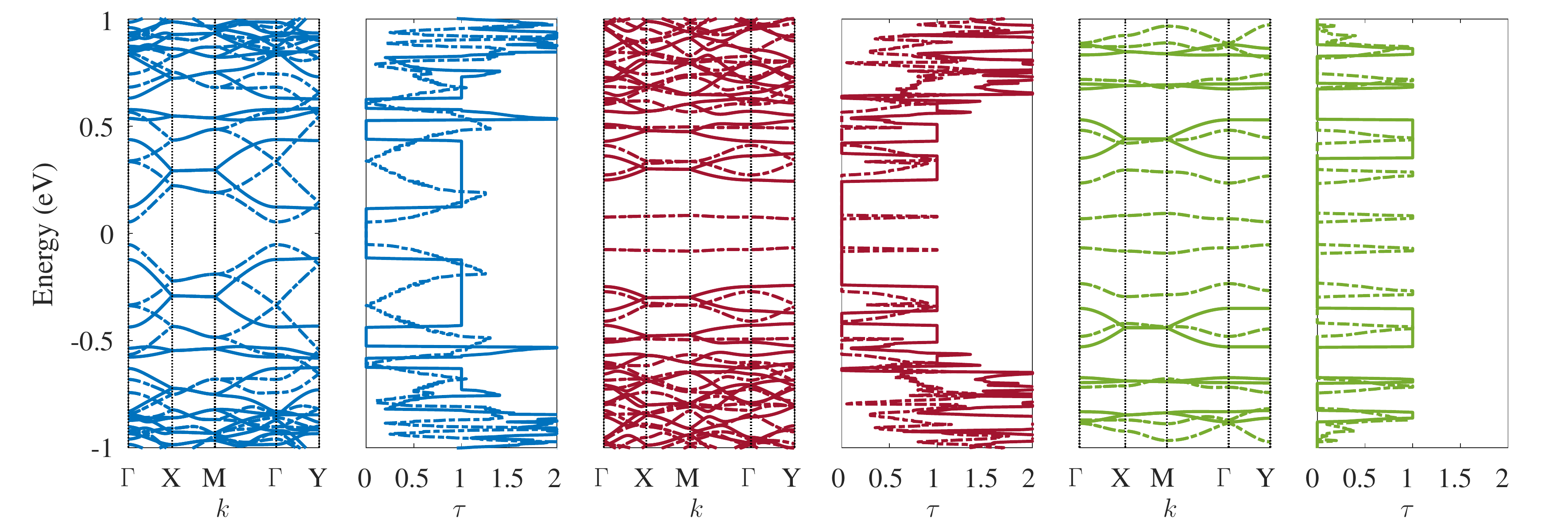}
	\caption{
		Band structures and transmission functions of $\alpha$ (blue), $\beta$
		(red), $\gamma$ (green) respectively. Solid (dash-dotted) curves represent the asymmetric (symmetric) structures. Zero of the energy corresponds to the charge neutrality point.
	}
	\label{fig:band_trans}
\end{figure*}

GALs can have high TE efficiencies~\cite{karamitaheri2011,gunst2011,chang2012,chang2014,sadeghi2015}.
It was shown that a few repetitions of the GAL unit cell are enough to open a band gap and enhance $zT$ significantly in two-dimensional graphene~\cite{gunst2011}.
In one-dimensional zigzag graphene nanoribbons, $zT$ was predicted to have very large values due to perforation with nanopores, where the electronic quality at the edges were preserved~\cite{chang2012,chang2014,sadeghi2015,yan:physletta:2012}.

In this letter, we consider three different GAL structures with symmetric and asymmetric geometries and show that breaking of symmetry in {2D} GAL geometry gives rise to a dimensional crossover in their electronic structure.
The dimensional crossover is characterized with nondispersive bands in the transverse direction, which is associated with a plateau in the transmission spectrum.
Namely, asymmetric GALs form {an array of one-dimensional electron systems placed parallel to each other} and show stepwise transmission spectrum for low energy carriers.
As a result, thermopower, power factor  and $zT$ are enhanced considerably.

\section{SYSTEMS AND METHODS}
In this work, we consider hexagonal antidots and rectangular unit cells without loss of generality. The parameters defining the GALs are shown in Fig.~\ref{fig:structure}a.
A unit cell contains two antidots.
$L_x$ and $L_y$ are the unit cell parameters (measured in terms of the carbon-carbon distance).
The origin is chosen at the center of an antidot, the other antidot having distances $\Delta x$ and $\Delta y$ from the origin.
The antidot size is $r$.
We consider three sizes for the unit cells (to be denoted as $\alpha$, $\beta$ and $\gamma$) with symmetric and asymmetric configurations. 
(i.e. $\alphaS$, $\alphaA$, etc.)
Parameters for the considered structures are tabulated in Table~\ref{table:structural_parameters}.

We use first nearest neighbor tight-binding approximation, {which accurately describes the GAL structures, agreeing well with density functional theory in electronic band structures around Fermi level~\cite{furst2009}},
\begin{align}
	H=\sum_i E_ic_i^\dagger c_i
	 -t\sum_{\langle i,j\rangle} 
	 (	 c_i^\dagger c_j + c_j^\dagger c_i  ).
\end{align}
Here $c_i$ ($c_i^\dagger$) is electron annihilation (creation) operator for the $p_z$ orbital of the $i$th atom, $E_i$ is the onsite energy, $t=2.7$~eV is the electron hopping energy, and $\langle i,j\rangle$ denotes the nearest neighbor atoms. 
$E_i=0$ for ballistic calculations and picks random values when Anderson disorder is introduced.

Transmission spectra are calculated using the nonequilibrium Green function (NEGF) methodology~\cite{datta:book:1997,ryndyk:book:2015}.
The system is partitioned as the left electrode, the central region and the right electrode ($L$, $C$ and $R$). Hence, $H=H_C+\sum_{\nu}(H_\nu+H_{C\nu}+H_{\nu C})$ with $\nu=L,R$.
For disordered systems, $E_i$ is nonzero only in the central region.
Periodic boundary conditions are employed in the transverse direction, 
{and the wave-vector in the transverse direction is denoted as $k_y$.}
The retarded Green function for the central region is defined as
$G(E,{k_y})=[E+i\delta-H_C({k_y})-\Sigma(E,{k_y})]^{-1}$, where $\delta$ is an infinitesimal positive number, and the self energy term contains contributions from the reservoirs, $\Sigma=\Sigma_L+\Sigma_R$.
The self energy due to a reservoir $\nu$ is obtained using
$\Sigma_{\nu}=H_{C\nu}g_{\nu}H_{\nu C}$, where $g_\nu=(E+i\delta-H_\nu)^{-1}$ is the free Green function for the reservoir.
Tranmission spectra are obtained using the Green functions as 
\begin{align}
	\tau(E)=
	\frac{L_y}{\pi}
	\int d{k_y}\,
	\mathrm{Tr}
	\left[
	\Gamma_L G \Gamma_R G^\dagger
	\right],
\end{align}
with $\Gamma_\nu=i(\Sigma_\nu-\Sigma_\nu^\dagger)$ being broadening matrices.
The $k$-space integration is carried out using 20 $k$-points.

\begin{table}[b]
	\caption{\label{table:structural_parameters} Structural parameters for $\alpha$, $\beta$ and $\gamma$ (normalized with carbon-carbon bond length, $d_\textrm{CC} = 1.42\textrm{\AA}$). Subscript $S$ ($A$) implies that the structure has symmetric (asymmetric) configuration. $L_x$ and $L_y$ are the unit cell width and height, $\Delta x$ and $\Delta y$ are the minimum relative distances between antidots in $x-$ and $y-$ directions, $r$ is the antidot size.}
	\begin{ruledtabular}
		\centering
		\begin{tabular}{ l c c c c c}
			& $L_x$ & $L_y$ & $\Delta x$ & $\Delta y$ & $r$ \\
			\hline
			$\alpha_A$ & 39 & 23$\sqrt{3}$ & 19 & 7.5  $\sqrt{3}$ & {9.5}  \\ 
			$\alpha_S$ & 39 & 23$\sqrt{3}$ & 19 & 11.5 $\sqrt{3}$ & {9.5}  \\ 
			$\beta_A$  & 48 & 27$\sqrt{3}$ & 24 & 10   $\sqrt{3}$ & {12.5} \\ 
			$\beta_S$  & 48 & 27$\sqrt{3}$ & 24 & 13.5 $\sqrt{3}$ & {12.5} \\ 
			$\gamma_A$ & 30 & 19$\sqrt{3}$ & 15 & 11   $\sqrt{3}$ & {10}   \\ 
			$\gamma_A$ & 30 & 19$\sqrt{3}$ & 15 & 9.5  $\sqrt{3}$ & {10}   \\ 
		\end{tabular}
	\end{ruledtabular}
\end{table}

\begin{figure}[b]
	\centering
	\includegraphics[width=8.5cm]{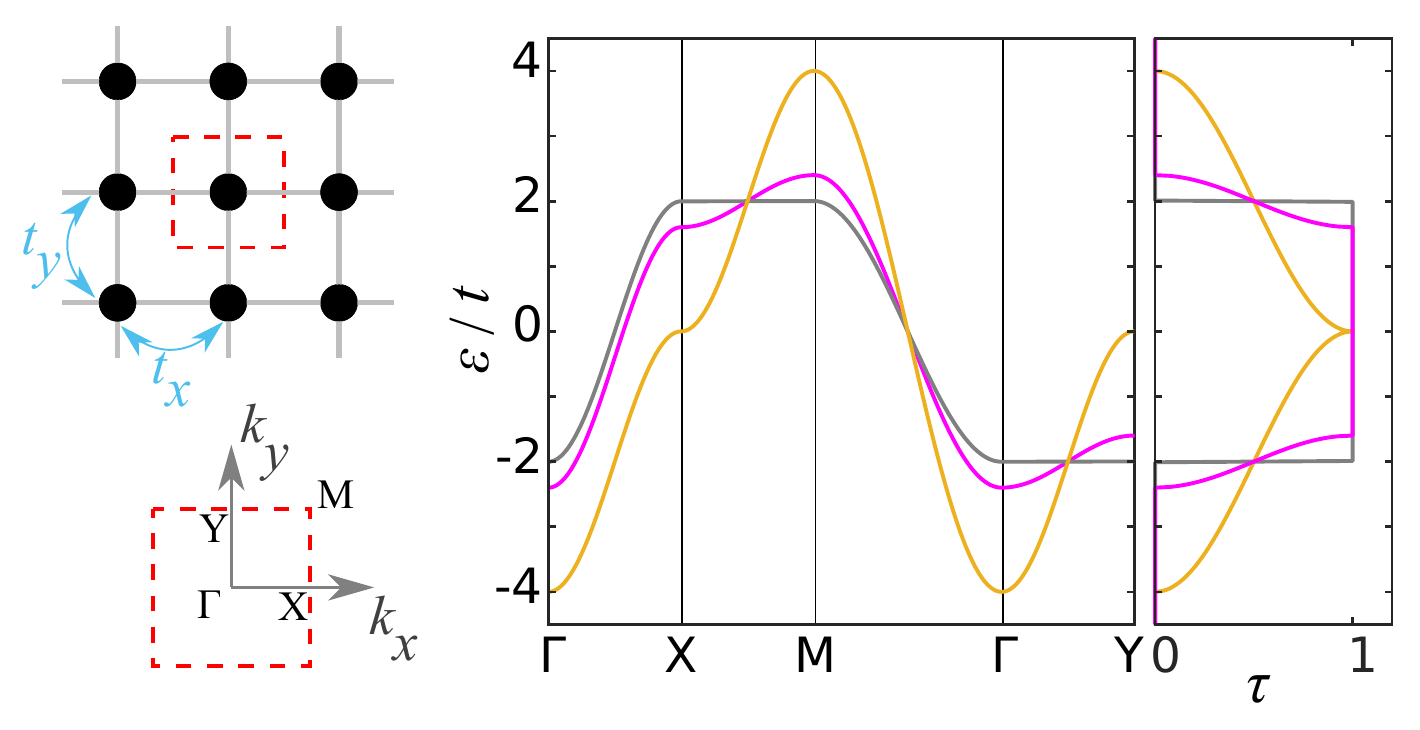}
	\vspace{-5mm}
	\caption{
		Toy model with a square lattice in the first nearest neighbor approximation having longitudinal and transverse hopping parameters $t_x$ and $t_y$.
		The electronic bands and corresponding transmission spectra are plotted for different values of $\eta=t_y/t_x$, namely  $\eta=1$, 0.2 and 0  (orange, magenta and gray, respectively).
	}
	\label{fig:toy_model}
\end{figure}

Thermoelectric figure of merit is defined as 
\begin{align}
   zT=\frac{S^2\sigma T}{\kappael+\kappaph},
\end{align}
where $S$ is the Seebeck coefficient, $\sigma$ is electrical conductance, $\kappael$ ($\kappaph$) is the electronic (phononic) contribution to thermal conductance, and $T$ is temperature.
The electronic coefficients are calculated using the integrals~\cite{sivan:physrev:1986}
\begin{align}
	L_n(\mu,T)=-\frac{2}{h}
	\int \frac{\partial f_{FD}}{\partial T}(E-\mu)^n \tau(E)\,dE,
\end{align}
as $S=(eT)^{-1}L_1/L_0$, $\sigma=e^2L_0$, and 
$\kappael=T^{-1}(L_2-L_1^2/L_0)$, where $f_{FD}(E,\mu,T)$ is the Fermi-Dirac distribution function and $\mu$ is the chemical potential.
Phononic contribution to thermal conductance is obtained using Green functions as~\cite{rego:prl:1998,sevincli:jpcm:2019}
\begin{align}
	\kappaph=\int 
	\frac{\partial f_{BE}}{\partial T}\,
	\hbar\omega\,
	\zeta(\omega)\,	
	\frac{d\omega}{2\pi},
	\label{eqn:kappa_ph}
\end{align}
$f_{BE}(\omega,T)$ being the Bose-Einstein distribution function, 
and $\zeta(\omega)$ stands for the phonon transmission spectrum.
Fourth-nearest-neighbor force constant approximation is used for constructing the dynamical matrices~\cite{saito:book:1998,zimmermann:prb:2008}.

\section{RESULTS AND DISCUSSION}
We first calculate the electronic structures and plot the band structures of symmetric and asymmetric systems for comparison. (see Fig.~\ref{fig:band_trans}-left panels)
For $\alpha$-GALs, the band dispersions are similar along $\Gamma$X and XM directions around the charge neutrality point (CNP).
Low energy bands are dispersive along M$\Gamma$ for both $\alphaS$ and $\alphaA$ structures.
However $\alphaA$ structure has nondispersive bands along $\Gamma$Y unlike $\alphaS$.
Such flat bands appear along $\Gamma$Y in $\betaA$ and $\gammaA$ structures too.
Symmetric structures $\betaS$ and $\gammaS$ have flat valence and conduction bands, which are dispersionless not only along $\Gamma$Y but in the entire Brillouin zone.
Plotting the modulus square of their wave-functions at the $\Gamma$ point, $|\psi_{\bold{k}}(\bold{x})|^2$, we observe that these flat band states in $\betaS$ and $\gammaS$ are localized states around the antidots due to dangling bonds and are not in the main focus of this study.
In asymmetric GAL structures, transmission spectra show plateaus with integer values. (Fig.~\ref{fig:band_trans})
The flat portions of the bands and transmission plateaus are related such that separation of flat bands along the $\Gamma$Y direction is the same with the widths of plateaus for all $\alphaA$, $\betaA$ and $\gammaA$ structures.

\begin{figure}[t]
	\includegraphics[width=85mm]{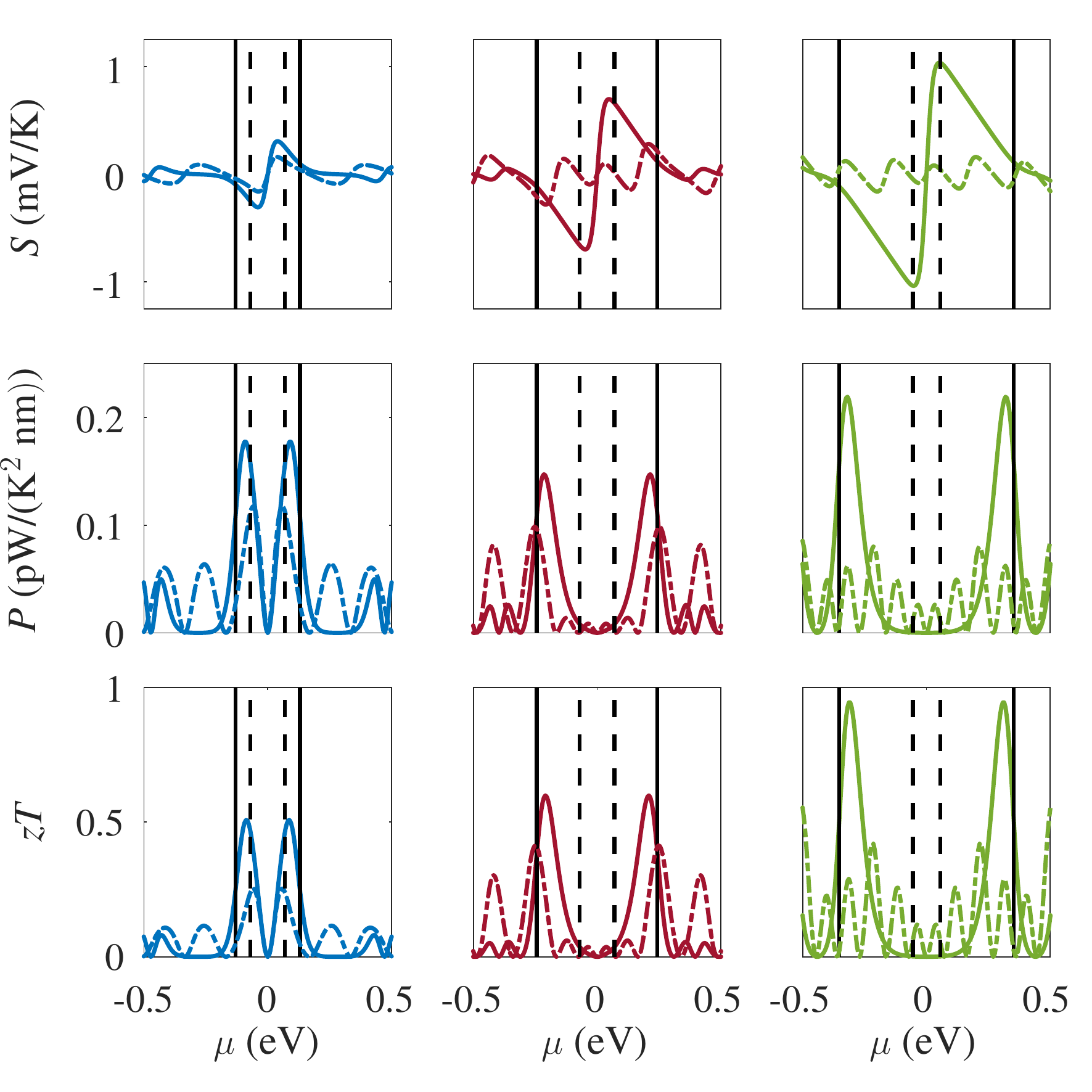}
	\caption{Thermoelectric properties. The columns represent $\alpha$, $\beta$ and $\gamma$ geometries, respectively, while the rows stand for the Seebeck coefficient ($S$), power factor ($P$), and TE efficient ($zT$). Values for symmetric (asymmeteric) structures are plotted using dot-dashed (solid) curves. Band gap edges are represented with vertical lines (dot-dashed) for asymmetric (symmetric) GALs.
		\label{fig:te}
	}
\end{figure}

The flat bands and the transmission plateaus stem from the symmetry breaking.
In order to explore this feature, we consider a toy model consisting of a square lattice in the first nearest neighbor approximation, with lattice constant $a$, and compare isotropic, anistropic ({effectively} 1D) and truly one-dimensional lattices. (see Fig.~\ref{fig:toy_model})
The band dispersion for such a system is 
$E(k_x,k_y)=-2t_x\cos\,k_x-2t_y\cos\,k_y$, 
where $t_x$ and $t_y$ are the hopping parameters in longitudinal and transverse directions, respectively.
Width normalized ballistic transmission along the $x-$direction can be obtained analytically as $\tau(\varepsilon)/a=\mathrm{Re}[\arccos\,\xi]/\pi$.
For the symmetric case ($t_x=t_y=t$), the square lattice has the well-known electronic band structure
and $\xi=\varepsilon/2-1$ (with $\varepsilon=|E/t|$).
That is, $\tau$ is a smoothly varying function except at $\varepsilon=0$, where it is  maximized.
For the asymmetric case ($t_y/t_x=\eta<1$), the band lies within $[-2(1+\eta),2(1+\eta)]$.
It is as dispersive as the symmetric one along $\Gamma$X,
but the dispersions along $\Gamma$Y (and XM) are limited to 4$\eta$.
As asymmetry increases (i.e. for smaller $\eta$) we have flatter bands in the transverse direction.
Correspondingly, transmission is altered with $\xi=(\varepsilon-2)/2\eta$.
The argument $|\xi|<1$ only within $\pm2-2\eta<\varepsilon<\pm2+2\eta$,
and $\xi<-1$ at the mid-band energies.
Hence there appears a transmission plateau,
$\tau(\varepsilon)/a=1$, for $-2+2\eta<\varepsilon<2-2\eta$ as shown with the magenta curve in Fig.~\ref{fig:toy_model}, where $\eta=0.2$ is chosen.
Anisotropic transmission spectrum with the plateau is a signature of dimensional crossover.
As $\eta\rightarrow0$, the square lattice is transformed to independent parallel linear chains, the band dispersion along $\Gamma$Y and XM vanishes, and $\xi\rightarrow-\infty$ inside $-2<\varepsilon<2$ to yield a stepwise transmission, a characteristic of one-dimensional systems. (gray curve in Fig.~\ref{fig:toy_model})
{In strictly 1D structures, stepwise transmission is a consequence of the fact that there is no transverse direction in which the energy disperses to give rise to changes in transmission values. In effectively {effectively 1D} structures, on the other hand, the energy dispersion is the transverse direction is very small, if not zero, as a result the transmission is not affected from the transverse component of the wave vector and the transmission is stepwise. }

The anisotropic square lattice model reveals the origins of nondispersive bands along the $\Gamma$Y direction of asymmetric GALs and the corresponding abrupt (or sometimes stepwise) changes in the transmission spectrum, namely the {effective} 1D character.
We note that not all the electronic bands of asymmetric GALs display an {effective} 1D behavior.
In $\alphaA$, $\betaA$ and $\gammaA$ structures, the highest valence band (the lowest conduction band) is {effectively} 1D, whereas there exist lower (higher) valence (conduction) bands that are dispersive.
We also note that, depending on the structural parameters, asymmetric GALs can also have dispersive frontier bands as well as 1D deep bands.

\begin{figure}[b]
	\includegraphics[width=85mm]{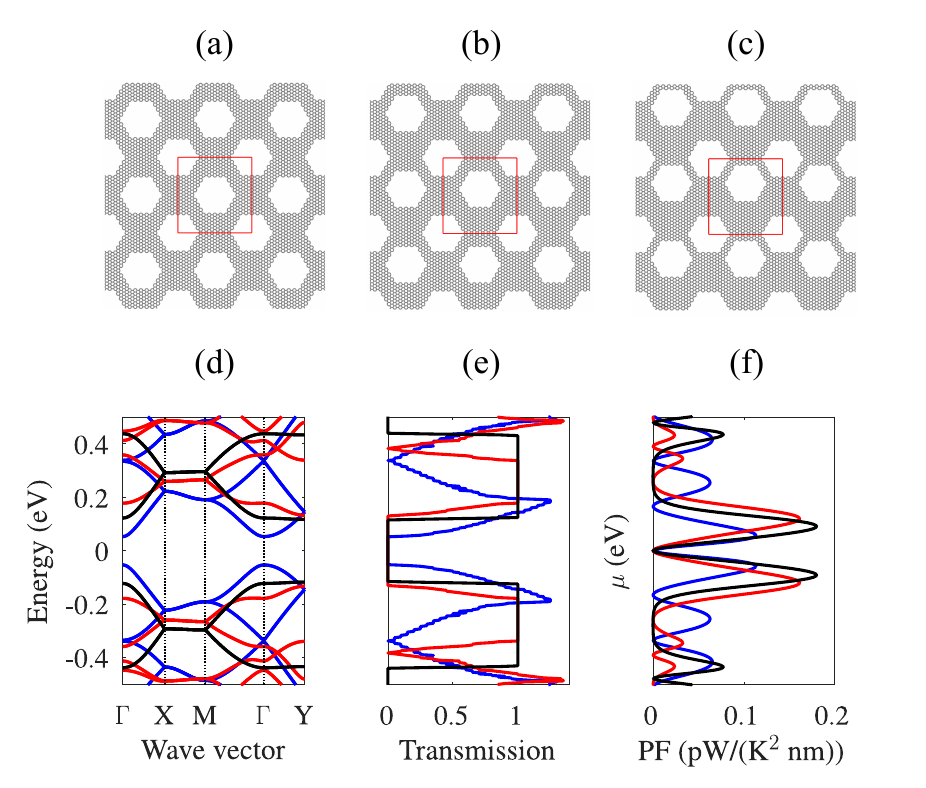}
	\caption{
		Three variations of $\alpha$ structure with their band structures, transmission functions and power factors are given (lower half). Blue, red and black lines correspond to the first ($\alphaS$), the second and the third ($\alphaA$) structure respectively.
		\label{fig:transition}
	}
\end{figure}

Next, we investigate the thermoelectric properties.
Fig.~\ref{fig:te} shows the Seebeck coefficient, power factor ($P=S^2\sigma$) and $zT$ calculated at $T=300$~K.
Dimensional crossover  in asymmetric structures enhance the Seebeck coefficient substantially.
This can be understood from the low temperature approximation~\cite{cutler:physrev:1969},
\begin{align}
	\label{eqn:seebeck}
	S\approx\frac{\pi^2\kB^2T}{3e}\frac{\partial\,\mathrm{ln}\,\tau(E)}{\partial E},
\end{align}
which shows that abrupt changes in the transmission spectrum maximize $S$.
In the first row of Fig.~\ref{fig:te}, one observes that $S$ is maximized inside the band gaps, where $\sigma$ is exponentially small. Therefore it is more convenient to compare the power factors. 
{Fig.~\ref{fig:te} shows that the maximum power factors are of the order $10^{-4} \textrm{W} \textrm{m}^{-1} \textrm{K}^{-2}$, which are comparable to the efficient thermoelectric graphene based devices in literature~\cite{Zong2020}.}
It is also observed that asymmetric structures have significantly larger $P$, compared to their symmetric counterparts.
$P$ is maximized for $\mu$ close to the band edges and the maximum $zT$ values are substantially larger for the asymmetric structures. 
The maximum $zT$ values achieved at the frontier band edges are
0.5, 0.6 and 0.95 (0.2, 0.4 and 0.4) at $T=300$~K for $\alphaA$, $\betaA$ and $\gammaA$ ($\alphaS$, $\betaS$ and $\gammaS$), respectively.
We should emphasize that, for asymmetric structures the maximum $zT$ values are achieved at chemical potentials close to the valence and conduction band edges, whereas they are obtained at second or deeper bands for $\betaS$ and $\gammaS$ geometries.

\begin{figure}[t]
	\includegraphics[width=75mm]{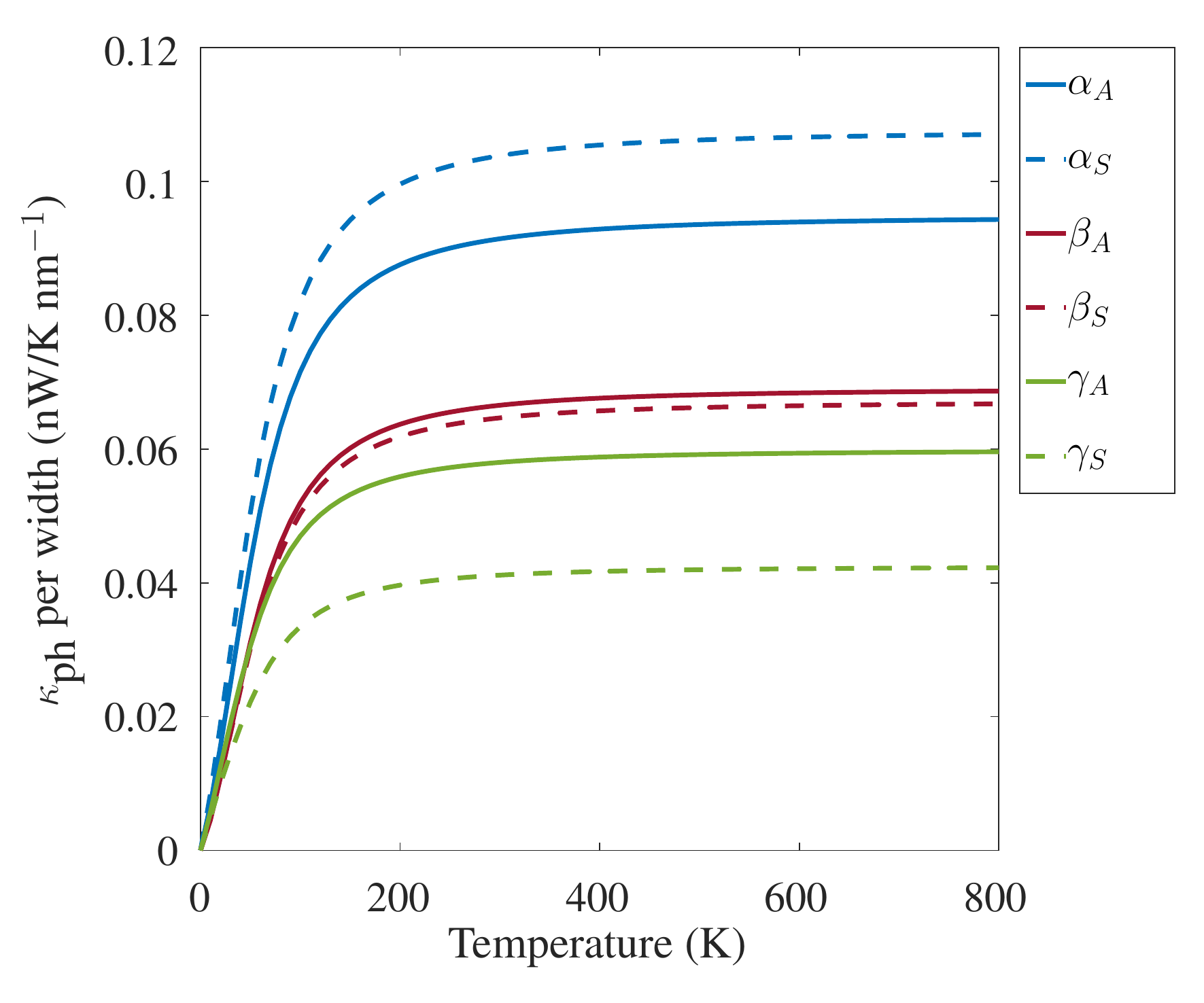}
	\caption{
		Width normalized phonon thermal conductance values as functions of temperature. Asymmetry changes phononic heat conduction for $\alpha$ and $\gamma$ whereas it remains almost unchanged for $\beta$.
		\label{fig:thermalconductance}
	}
\end{figure}

{
The crossover from {2D} to {effective 1D} can be demonstrated by examining three variations of the $\alpha$-structure, including $\alphaA$ and $\alphaS$. 
In Fig.~\ref{fig:transition}a,
the structures change from the symmetric configuration (Fig.~\ref{fig:transition}a) as the antidot in the middle of the unit cell is shifted upwards. The amount of shift is one benzene ring in each case (Fig.~\ref{fig:transition}b and \ref{fig:transition}c). 
When asymmetry is introduced, crossover to {effective 1D} behavior starts.
The bands along the $\Gamma$Y direction become less dispersive, accordingly a tranmission plateau occurs.
When asymeetry increases, i.e. $\alphaA$, the frontier bands become totally flat along $\Gamma$Y and transmission becomes stepwise.
Correspondingly, the Seebeck coefficient and hence the power factor ($P=S^2G$) increases with asymmetry. (see Eqn.~\ref{eqn:seebeck})
}

\begin{figure}[t]
	\includegraphics[width=75mm]{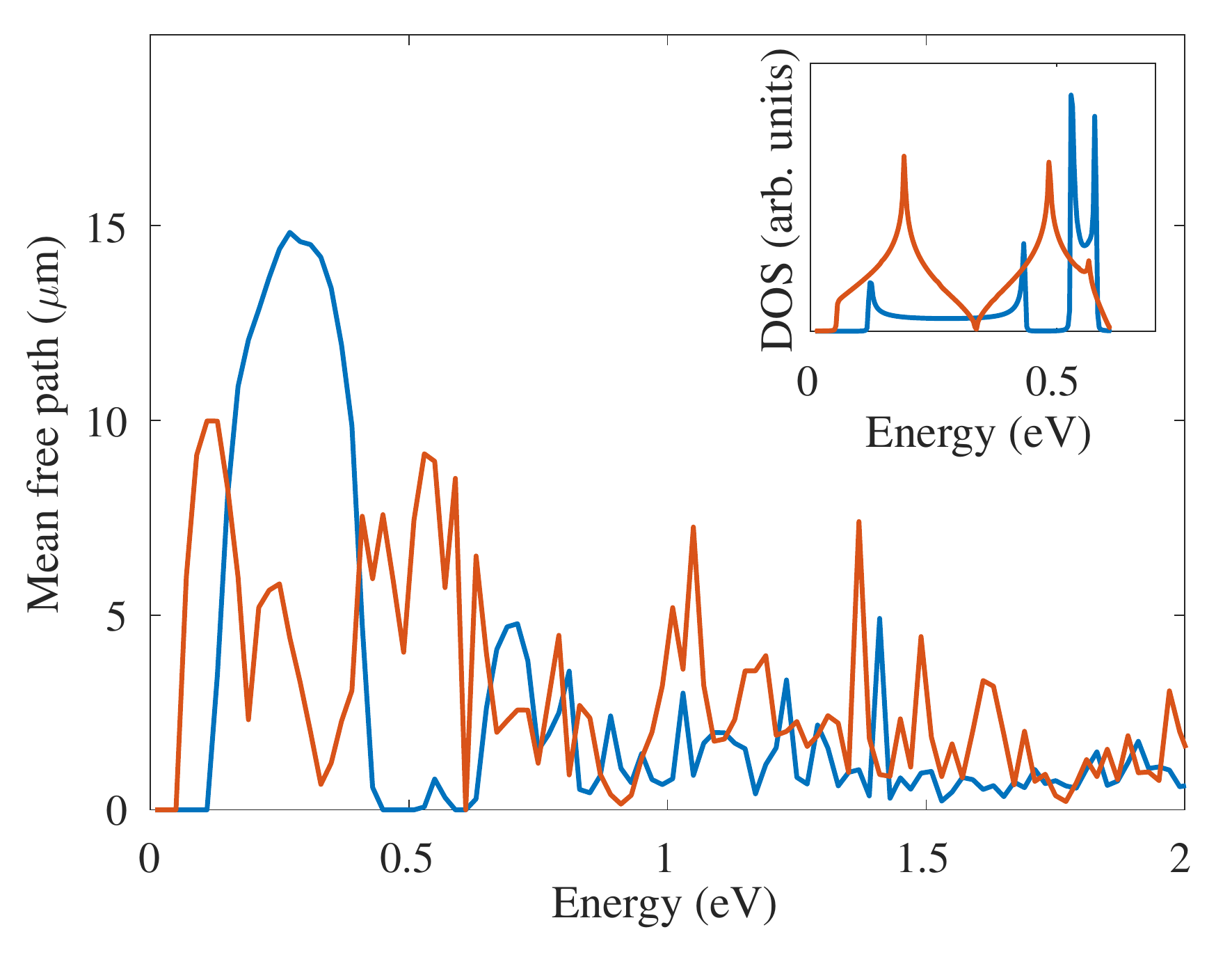}
	\caption{
		Mean free paths for $\alphaA$ and $\alphaS$ structures. Electronic density of states is shown in the inset.
		{Red curves represent the symmetric structure, whereas the blue curves are for the asymmetric ones.}
		\label{fig:mfp}
	}
\end{figure}

A major factor that determines $zT$ is the phonon contribution the thermal conductance, which
is  highest for $\alpha$ and lowest for $\gamma$ geometries (Fig.~\ref{fig:thermalconductance}).
This order is in agreement with the ordering of geometries with respect to the ratio between area of antidot to that of the unit cell.
Namely, the areal density of the structure is the key ingredient for determining its phonon thermal conductance.
This finding is in agreement with previous studies~\cite{gunst2011}.
Asymmetry affects phonon transmission, as well.
There appears transmission plateaus in asymmetric structures but their effect on heat transport is less decisive compared to electrons, which is a result of bosonic statistics. 
The weight factor $\omega\,\partial f/\partial T$  (cf. Eqn.~\ref{eqn:kappa_ph}) incorporates all frequencies equally at high temperatures, therefore swaps away the fine details of the phonon transmission spectrum. 
As a result, asymmetry reduces $\kappaph$ significantly, for $\alphaA$, giving rise to better enhancement of TE efficiency, but $\kappaph$ of $\beta$ geometries are almost insensitive to symmetry breaking.
On the other hand, $\kappaph$ of $\gammaS$ is lower than that of $\gammaA$.
When temperature increases from 300~K to 500~K, $\zTmax$ of asymmetric structures increase, whereas those of symmetric ones decrease. (see Table~\ref{table:ztmax}) This is due to narrower band gaps of symmetric structures, at which electron and hole contributions start to cancel each other for $E_\mathrm{gap}{<}10\kB T$.
When $T$ is further increased to 800~K, only $\gammaA$ has increased $\zTmax$, which has the widest band gap, and reaches the value of 1.76.

\begin{table}[b]
	\caption{The maximum $zT$ values for the studied structurres at various temperatures.}
	\label{table:ztmax}
	\begin{ruledtabular}
		\begin{tabular}{ccccccc}
			     & $\alphaS$ & $\alphaA$ & $\betaS$ & $\betaA$ & $\gammaS$ & $\gammaA$ 
			\\ \hline
			300 K & 0.25 & 0.51 & 0.41 & 0.60 & 0.58 & 0.94 \\
			500 K & 0.16 & 0.47 & 0.41 & 0.83 & 0.25 & 1.40 \\
			800 K & 0.08 & 0.22 & 0.28 & 0.79 & 0.17 & 1.76
		\end{tabular}
	\end{ruledtabular}
\end{table}

We further investigate electronic transmission in presence of Anderson type disorder with a uniform distribution of random onsite energies $E_i$ within $[-W/2,W/2]$, where $W=\sqrt{12}\kB T=90$~meV ($T=300$~K) is used for ensembles consisting of 100 samples.
The mean free path can be obtained by using transmission spectra of an ensemble of disordered systems, where
\begin{align}
	\frac{\langle\tau(E,L)\rangle}{\tau_0(E)}=
	\frac{\ell(E)}{\ell(E)+\lambda L},
\end{align}
with $\langle\tau(E,L)\rangle$ being ensemble averaged transmission, $\tau_0$ the pristine transmission, $L$ the length of the sample, $\ell$ the mean free path and $\lambda$ a constant depending on the dimensionality. 
Here, $\lambda=\pi/2$ ($\lambda=1$) is used for $\alphaS$ ($\alphaA$), which correspond to two-dimensional isotropic (strictly one-dimensional) systems~\cite{jeong:jap:2010}.
Fig.~\ref{fig:mfp} shows that  the mean free paths of {effectvely 1D} electrons at the transmission plateau are much longer than those of the symmetric configuration.
This is another consequence of dimensional crossover. 
{
The distribution of mean free path of effectively 1D electrons (blue curve within the interval 0.1~eV to 0.4~eV) is very similar to the typical strictly one-dimensional systems like carbon nanotubes. (see e.g. Ref.~\onlinecite{troizon:prb:2004})
}
The longer mean free paths are because of the flattened density of states of {effectively 1D} electrons.
According to Fermi's golden rule, the inverse scattering rate is proportional with the density of states.
For $\alphaS$, DOS is much higher than that of $\alphaA$. The {2D} DOS of $\alphaS$ has a peak at around 0.2~eV, giving rise to a dip in the mean free path.
The {effectively 1D} DOS has van Hove singularities at the band edges, a signature of one-dimensional systems, and a lower DOS value inside the transmission plateau.
We calculate TE coefficients in the presence of disorder using the ensemble averaged transmission values. Using the pristine phonon transmission values, we find a $\zTmax$ value of 0.35 for a 5~$\mu$m long $\alphaA$ structure at $T=300$~K. This value is still larger than that of the pristine symmetric structure.

\section{CONCLUSIONS}
We have shown that symmetry breaking can induce dimensional crossover in graphene antidot lattices and that it is possible to create parallel one-dimensional electron channels in series on a two-dimensional structure.
{
Although it is difficult to fabricate such symmetric or asymmetric GAL structures using top-down approaches, bottom-up fabrication techniques similar to those in Ref.~\onlinecite{moreno2018} can make it possible to realize such geometries.}
Dimensional crossover causes transmission plateaus, at whose edges thermopower is strongly enhanced, and $zT$ is increased.
Also, effectively one-dimensional electrons have much longer mean free paths than those which extend at two-dimensions.
{
Lastly, the strategy is not limited to graphene but should be possible to use in different 2D materials as well.}

\section*{ACKNOWLEDGEMENTS}
We acknowledge financial support from the Flag-Era JTC 2017 project 'MECHANIC' (funded by T\" UB\. ITAK
under Grant No. 117F480).

\bibliography{REFERENCES}

\end{document}